\documentclass[tighten, times, twocolumn]{aastex62}  % ApJ look-alike
\newcommand\apjcls{1}
\newcommand\aastexcls{2}
\newcommand\othercls{3}

\newcommand\papercls{\aastexcls}
\if\papercls \apjcls
\usepackage{apjfonts}
\else\if\papercls \othercls
\usepackage{epsfig}
\usepackage{margin}
\usepackage{times}
\fi\fi
\usepackage{ifthen}
\usepackage{natbib}
\usepackage{amssymb, amsmath}
\usepackage{appendix}
\usepackage{etoolbox}
\usepackage[T1]{fontenc}
\usepackage{paralist}

\if\papercls \apjcls
\newcommand\aas{\ref@jnl{AAS Meeting Abstracts}}% *** added by jh
\newcommand\dps{\ref@jnl{AAS/DPS Meeting Abstracts}}% *** added by jh
\newcommand\maps{\ref@jnl{MAPS}}% *** added by jh
\else\if\papercls \othercls
\usepackage{astjnlabbrev-jh}
\fi\fi

\if\papercls \aastexcls
\hypersetup{citecolor=blue, % color for \cite{...} links
            linkcolor=blue, % color for \ref{...} links
            menucolor=blue, % color for Acrobat menu buttons
            urlcolor=blue}  % color for \url{...} links
\else
\usepackage[%pdftex,      % hyper-references for pdflatex
bookmarks=true,           %%% generate bookmarks ...
bookmarksnumbered=true,   %%% ... with numbers
colorlinks=true,          % links are colored
citecolor=blue,           % color for \cite{...} links
linkcolor=blue,           % color for \ref{...} links
menucolor=blue,           % color for Acrobat menu buttons
urlcolor=blue,            % color for \url{...} links
linkbordercolor={0 0 1},  %%% blue frames around links
pdfborder={0 0 1},
frenchlinks=true]{hyperref}
\fi

\if\papercls \othercls

\else

\fi

\providecommand{\adsurl}[1]{\href{#1}{ADS}}

\makeatletter
\patchcmd{\NAT@citex}
  {\@citea\NAT@hyper@{%
     \NAT@nmfmt{\NAT@nm}%
     \hyper@natlinkbreak{\NAT@aysep\NAT@spacechar}{\@citeb\@extra@b@citeb}%
     \NAT@date}}
  {\@citea\NAT@nmfmt{\NAT@nm}%
   \NAT@aysep\NAT@spacechar\NAT@hyper@{\NAT@date}}{}{}

\patchcmd{\NAT@citex}
  {\@citea\NAT@hyper@{%
     \NAT@nmfmt{\NAT@nm}%
     \hyper@natlinkbreak{\NAT@spacechar\NAT@@open\if*#1*\else#1\NAT@spacechar\fi}%
       {\@citeb\@extra@b@citeb}%
     \NAT@date}}
  {\@citea\NAT@nmfmt{\NAT@nm}%
   \NAT@spacechar\NAT@@open\if*#1*\else#1\NAT@spacechar\fi\NAT@hyper@{\NAT@date}}
  {}{}
\makeatother

\makeatletter
\DeclareRobustCommand{\lowcase}[1]{\@lowcase#1\@nil}
\def\@lowcase#1\@nil{\if\relax#1\relax\else\MakeLowercase{#1}\fi}
\makeatother

\DeclareSymbolFont{UPM}{U}{eur}{m}{n}
\DeclareMathSymbol{\umu}{0}{UPM}{"16}
\let\oldumu=\umu
\renewcommand\umu{\ifmmode\oldumu\else\math{\oldumu}\fi}

\if\papercls \othercls

\else

\fi

\let\oldsim=\sim
\renewcommand\sim{\ifmmode\oldsim\else\math{\oldsim}\fi}
\let\oldpm=\pm
\renewcommand\pm{\ifmmode\oldpm\else\math{\oldpm}\fi}
\newcommand\by{\ifmmode\times\else\math{\times}\fi}
\newbox{\wdbox}
\renewcommand\c{\setbox\wdbox=\hbox{,}\hspace{\wd\wdbox}}
\renewcommand\i{\setbox\wdbox=\hbox{i}\hspace{\wd\wdbox}}

\newcount\timect
\newcount\hourct
\newcount\minct
\newcommand\now{\timect=\time \divide\timect by 60
         \hourct=\timect \multiply\hourct by 60
         \minct=\time \advance\minct by -\hourct
         \number\timect:\ifnum \minct < 10 0\fi\number\minct}

\catcode`@=11

\newcommand\comment[1]{}

\newcommand\commenton{\catcode`\%=14}

\renewcommand\math[1]{$#1$}
\newcommand\mathshifton{\catcode`\$=3}

\let\atab=&
\newcommand\atabon{\catcode`\&=4}

\let\oldmsp=\sp
\let\oldmsb=\sb
\def\sp#1{\ifmmode
           \oldmsp{#1}%
         \else\strut\raise.85ex\hbox{\scriptsize #1}\fi}
\def\sb#1{\ifmmode
           \oldmsb{#1}%
         \else\strut\raise-.54ex\hbox{\scriptsize #1}\fi}
\newbox\@sp
\newbox\@sb
\def\sbp#1#2{\ifmmode%
           \oldmsb{#1}\oldmsp{#2}%
         \else
           \setbox\@sb=\hbox{\sb{#1}}%
           \setbox\@sp=\hbox{\sp{#2}}%
           \rlap{\copy\@sb}\copy\@sp
           \ifdim \wd\@sb >\wd\@sp
             \hskip -\wd\@sp \hskip \wd\@sb
           \fi
        \fi}
\def\msp#1{\ifmmode
           \oldmsp{#1}
         \else \math{\oldmsp{#1}}\fi}
\def\msb#1{\ifmmode
           \oldmsb{#1}
         \else \math{\oldmsb{#1}}\fi}

\def\supon{\catcode`\^=7}

\def\subon{\catcode`\_=8}

\def\supsubon{\supon \subon}

\newcommand\actcharon{\catcode`\~=13}

\newcommand\paramon{\catcode`\#=6}

\comment{And now to turn us totally on and off...}

\newcommand\reservedcharson{ \commenton  \mathshifton  \atabon  \supsubon 
                             \actcharon  \paramon}

\catcode`@=12
\reservedcharson

\if\papercls \apjcls

\else

\fi

\newcommand\JWST{{\em JWST}}

\newcommand\chisq{\ifmmode{\chi\sp{2}}\else\math{\chi\sp{2}}\fi}
\newcommand\redchisq{\ifmmode{ \chi\sp{2}\sb{\rm red}}
                    \else\math{\chi\sp{2}\sb{\rm red}}\fi}
\newcommand\Teq{\ifmmode{T\sb{\rm eq}}\else$T$\sb{eq}\fi}
\newcommand\Teff{\ifmmode{T\sb{\rm eff}}\else$T$\sb{eff}\fi}
\newcommand\mjup{\ifmmode{M\sb{\rm Jup}}\else$M$\sb{Jup}\fi}
\newcommand\rjup{\ifmmode{R\sb{\rm Jup}}\else$R$\sb{Jup}\fi}
\newcommand\msun{\ifmmode{M\sb{\odot}}\else$M\sb{\odot}$\fi}
\newcommand\rsun{\ifmmode{R\sb{\odot}}\else$R\sb{\odot}$\fi}
\newcommand\mearth{\ifmmode{M\sb{\oplus}}\else$M\sb{\oplus}$\fi}
\newcommand\rearth{\ifmmode{R\sb{\oplus}}\else$R\sb{\oplus}$\fi}
\newcommand\moloxy{O$\sb{2}$}
\newcommand\molnit{N$\sb{2}$}
\newcommand\ozone{O$\sb{3}$}

\newcommand\methane{CH$\sb{4}$}
\newcommand\methyl{CH$\sb{3}$}
\newcommand\water{H$\sb{2}$O}
\newcommand\carbdiox{CO$\sb{2}$}
\newcommand\carbmono{CO}
\newcommand\ethane{C$\sb{2}$H$\sb{6}$}
\shorttitle{M dwarf Stellar Effects on Habitable Atmospheres}
\shortauthors{Afrin Badhan {\em et al.}}
\usepackage{amsmath}
\usepackage{textcomp}

\usepackage{graphicx}
\usepackage[caption=false]{subfig}

\begin{document}
\title{\vspace{-0.4cm}\hbox{Stellar Activity Effects on Moist Habitable Terrestrial Atmospheres Around M dwarfs}}
%% AUTHOR/INSTITUTIONS FOR AASTEX6.1:
\author{Mahmuda~Afrin~Badhan}
\affiliation{NASA Goddard Space Flight Center, 8800 Greenbelt Rd, Mail Stop 699.0 Building 34, Greenbelt, MD 20771, USA}
\affiliation{Department of Astronomy, University of Maryland College Park, College Park, MD, USA}
\affiliation{NASA Astrobiology Institute's Virtual Planetary Laboratory, P.O. Box 351580, Seattle, WA 98195, USA}
\affiliation{Sellers Exoplanet Environments Collaboration, NASA Goddard Space Flight Center, Greenbelt, MD 20771, USA}

\author{Eric~T.~Wolf}
\affiliation{NASA Astrobiology Institute's Virtual Planetary Laboratory, P.O. Box 351580, Seattle, WA 98195, USA}
\affiliation{Sellers Exoplanet Environments Collaboration, NASA Goddard Space Flight Center, Greenbelt, MD 20771, USA}
\affiliation{Department of Atmospheric and Oceanic Sciences, Laboratory for Atmospheric and Space Physics, University of Colorado Boulder, Boulder, CO 80309, USA}

\author{Ravi~Kumar~Kopparapu}
\affiliation{NASA Goddard Space Flight Center, 8800 Greenbelt Rd, Mail Stop 699.0 Building 34, Greenbelt, MD 20771, USA}
\affiliation{NASA Astrobiology Institute's Virtual Planetary Laboratory, P.O. Box 351580, Seattle, WA 98195, USA}
\affiliation{Sellers Exoplanet Environments Collaboration, NASA Goddard Space Flight Center, Greenbelt, MD 20771, USA}

\author{Giada~Arney}
\affiliation{NASA Goddard Space Flight Center, 8800 Greenbelt Rd, Mail Stop 699.0 Building 34, Greenbelt, MD 20771, USA}
\affiliation{NASA Astrobiology Institute's Virtual Planetary Laboratory, P.O. Box 351580, Seattle, WA 98195, USA}
\affiliation{Sellers Exoplanet Environments Collaboration, NASA Goddard Space Flight Center, Greenbelt, MD 20771, USA}

\author{Eliza M.-R. Kempton}
\affiliation{Department of Astronomy, University of Maryland College Park, College Park, MD, USA}
\affiliation{Department of Physics, Grinnell College, Grinnell, IA 50112, USA}

\author{Drake~Deming}
\affiliation{Department of Astronomy, University of Maryland College Park, College Park, MD, USA}
\affiliation{NASA Astrobiology Institute's Virtual Planetary Laboratory, P.O. Box 351580, Seattle, WA 98195, USA}

\author{Shawn~D.~Domagal-Goldman}
\affiliation{NASA Goddard Space Flight Center, 8800 Greenbelt Rd, Mail Stop 699.0 Building 34, Greenbelt, MD 20771, USA}
\affiliation{NASA Astrobiology Institute's Virtual Planetary Laboratory, P.O. Box 351580, Seattle, WA 98195, USA}
\affiliation{Sellers Exoplanet Environments Collaboration, NASA Goddard Space Flight Center, Greenbelt, MD 20771, USA}

%% AUTHOR/INSTITUTIONS FOR EMULATE APJ:
% \author{Patricio~E.~Cubillos\altaffilmark{1,2},
% Joseph~Harrington\altaffilmark{1},
% and
% Third~Author\altaffilmark{1}
% }
% \affil{\sp{1} Planetary Sciences Group, Department of
%               Physics, University of Central Florida, Orlando, FL 32816-2385\\
%        \sp{2} Space Research Institute, Austrian Academy of Sciences,
%               Schmiedlstrasse 6, A-8042, Graz, Austria}

\email{afrin20m@astro.umd.edu}

 %% Extra info for aastex:
 %\received{{Yesterday}}
 %\revised{Today}
 %\accepted{Tonight}
 %\published{Tomorrow}
 %\submitjournal{AASJournal}

\begin{abstract}

Transit spectroscopy of terrestrial planets around nearby M dwarfs is a primary goal of space missions in coming decades. 3-D climate modeling has shown that slow-synchronous rotating terrestrial planets may develop thick clouds at the substellar point, increasing the albedo. For M dwarfs with \Teff{} > 3000 K, such planets at the inner habitable zone (IHZ) have been shown to retain moist greenhouse conditions, with enhanced stratospheric water vapor ($fH_2O$ > 10\textsuperscript{-3}) and low Earth-like surface temperatures. However, M dwarfs also possess strong UV activity, which may effectively photolyze stratospheric \water{}. Prior modeling efforts have not included the impact of high stellar UV activity on the \water{}. Here, we employ a 1-D photochemical model with varied stellar UV, to assess whether \water{} destruction driven by high stellar UV would affect its detectability in transmission spectroscopy. Temperature and water vapor profiles are taken from published 3-D climate model simulations for an IHZ Earth-sized planet around a 3300 K M dwarf with an \molnit{}-\water{} atmosphere; they serve as self-consistent input profiles for the 1-D model. We explore additional chemical complexity within the 1-D model by introducing other species into the atmosphere.  We find that as long as the atmosphere is well-mixed up to 1 mbar, UV activity appears to not impact detectability of \water{} in the transmission spectrum. The strongest \water{} features occur in the \JWST{} MIRI instrument wavelength range and are comparable to the estimated systematic noise floor of \texttt{\char`\~}50 ppm. %We also find that even the highest UV scenario does not produce spectrally significant \ozone{}.
\end{abstract}
\vspace{-0.2cm}
% http://journals.aas.org/authors/keywords2013.html

\section{Introduction}
\label{introduction}
Planets orbiting close enough to their host M dwarf stars to be tidally-locked have their day-sides continuously subjected to the enormous UV stellar irradiation. The first habitable zone (HZ) exoplanets to have their atmospheres characterized will likely be such tidally-locked planets orbiting nearby M dwarf stars. Thus there is a need to understand the behavior of such planetary atmospheres. Observed spectroscopic signatures from transit measurements can reveal spectrally active species in a planet's atmosphere. Present observational technologies not only inform us about the observed atmosphere's composition, but can also shed light on the planet's physical properties and atmospheric dynamics. NASA's upcoming space missions, such as the \emph{James Webb Space Telescope (JWST)}, will meet the sensitivity and broad wavelength coverage requirements to constrain extrasolar planetary atmospheric composition with unprecedented accuracy. 

Chemical disequilibrium processes (e.g. photolysis, vertical convection, life, etc.) can be diagnosed in planetary atmospheres by studying the observed trends in the abundances of detected species (e.g. \citealt{2013bApJLine}). In the deep atmosphere, where pressures and temperatures are high, reaction timescales are short. So, species tend to stay in chemical equilibrium. Temperatures tend to decrease with increasing altitude in planetary atmospheres without thermal inversions, slowing the reaction rates to a point where vertical transport starts dominating, causing species to spread out. For slowly rotating Earth-like planets, changes to the large-scale circulation lead to an increased efficiency of vertical mixing \citep{2013APJYang}. This means the lower unobserved region--the troposphere--is able to communicate with the upper regions probed by our space infra-red (IR) instruments. In the uppermost regions, the high ultra-violet (UV) instellation (i.e. stellar insolation from stars other than the Sun) can lead to photolysis and increasing depletion in the abundances of some molecular species. Planets with equilibrium temperatures below 1200K (i.e. they receive < 340-400 times the Earth-equivalent instellation \emph{S\textsubscript{o}}, using geometric albedo range 0.01-0.15 from \citet{2013ApJHeng}), have been shown to have the most obvious signs of disequilibrium via chemical kinetics modeling \citep{2003ApJLiang,2004ApJLiang, 2009Zahnlea,2009Zahnleb, 2013bApJLine, 2011ApJMoses, 2013ApJMoses,2011ApJV&M,2012APJRKK,2012ApHu,2012ApJMREK}.

Disequilibrium mechanisms play a noticeable role in altering atmospheric composition at altitudes probed by remote sensing techniques. Disequilibrium sources within the solar system include Venus's sulfuric acid hazes (e.g. \citealt{1982IcarYung,1994IcarKrasnopolsky}), Titan's hydrocarbon hazes (e.g. \citealt{1980ApJAllen,1984ApJYung}, and even Earth's ozone--a product of O\textsubscript{2} photolysis (e.g. \citealt{1930aChapman,1930bChapman,1942Chapman}. Observational \citep{2014NatKnutson, 2014NatKreidberg, 2018ApJKreidberg, 2016ApJGreen,2015A&AWH,2017MNRASWH} and theoretical \citep{2013ApJB&S,2016ApJMbarek, 2017ApJArney,2013ApJHeng,2017AJMorley} studies have shown that clouds and hazes dominate both transmission and reflection spectra of all kinds of planets. They can alter the thermal structure and composition in the higher altitudes, in addition to masking spectral features from the lower atmosphere. 3-D climate simulations have shown that for synchronously rotating planets, persistent substellar clouds may act in favor of habitability by increasing planetary albedo and decreasing the surface temperature, allowing planets to remain habitable at higher stellar fluxes than that of Earth's (e.g. \citealt{2013APJYang,2016GRLWay,2016APJRKK,2017APJRKK,2018EPSLBin}). A synchronous orbit around the host star is a valid assumption for planets in the habitable zones of such low mass stars \citep{2015SciLeconte}.
%Observational and theoretical studies have shown that clouds and hazes dominate both transmission and reflection spectra of all kinds of planets \citep{2013ApJB&S, 2014NatKnutson, 2014NatKreidberg, 2018ApJKreidberg, 2016ApJGreen, 2016ApJMbarek, 2017ApJArney, 2013ApJHeng, 2017AJMorley, 2015A&AWH,2017MNRASWH}.

Planet rotation rate, and thus the Coriolis effect, plays a key role in modulating atmospheric circulation and climate (e.g. \citealt{2010JAMESMerlis,2014ApJYang, 2017IcarNoda, 2018AsBioFujii}).  For slowly rotating planets, the Coriolis effect is weak, and planets maintain large-scale day-night thermal circulation patterns. Such worlds are characterized by strong convection and upwelling air around their substellar point, and downwelling air on the antistellar side. These effects combine to yield strong vertical mixing of H\textsubscript{2}O, which creates the ubiquitous subtellar cloud deck for slow rotators (e.g. \citealt{2013APJYang}), and significantly enhances stratospheric H\textsubscript{2}O planet-wide \citep{2017APJRKK,2017ApJFujii}. Planetary stratospheres are more readily sensed by transit observations than tropospheres due to refraction effects (e.g. \citealt{2014ApJMisra}, \citealt{2015MNRASBetremieux}), and cloud opacity can prohibit observations of the relatively wet lower layers. The exact range of pressures accessible to and sensed by IR observations depends on the star-planet distance as well as the star and planet type, since the location of condensates vary with those parameters. \citet{2017APJRKK} found that slow rotators around M dwarfs maintain moist-greenhouse conditions (stratospheric H\textsubscript{2}O content >10\textsuperscript{-3}, \citealt{1993IcarKasting}), despite relatively mild surface temperatures (\texttt{\char`\~}280 K). Therefore, we may expect to observe stronger H\textsubscript{2}O features in the transmission spectra of habitable slow rotators around M dwarfs compared to a true Earth-twin; Earth has a relatively dry stratosphere (\emph{f\water{}}\texttt{\char`\~}10\textsuperscript{-6}). At 1 mbar (the model top of the 3-D climate model), vertical mixing should remain strong enough to compete with the photochemical H\textsubscript{2}O loss \citep{2014ApJYang,2016APJRKK,2017ApJFujii}. The question then becomes, is the H\textsubscript{2}O loss above 1 mbar from photodisassociation significant enough to affect our ability to detect it with \JWST{}? If so, can we quantify this effect?

To answer this, we study the composition of such planets with a 1-D atmospheric model that includes chemical kinetics (including photolysis) and vertical mixing. We explore the influence of UV irradiation on the composition of the atmosphere at the planet's terminator (simulated by a 3-D climate model), which is probed by transit transmission observations. We consider a 1-D model column extending above the stratosphere to explore chemical complexity from photochemical disequilibrium and its impact on observing. We look for any discernible impact on the spectra of a selected planet-star pair within the moist greenhouse regime of \citet{2017APJRKK}. Here we report our findings for an \molnit{}-\water{}-dominated moist terrestrial IHZ planet modeled around a 3300 K M dwarf with synthetically varied stellar UV emissions from 1216 {\AA} - 4000 {\AA}. We compare our spectra with \citet{2017APJRKK} and discuss implications for future observations of moist habitable atmospheres with \JWST{}. 

The rest of this paper is laid out as follows: In Section \ref{sec:methods}, we give an overview of the analysis method and describe the various modeling tools employed in this study. In Section \ref{sec:results}, we present our findings for each stellar profile scenario, for a total of five scenarios. In Section \ref{sec:discussion}, we discuss the implications of the results on molecular detection via future observations with \JWST{}, and we conclude in Section \ref{sec:conclusions}.

\section{Methods}
\label{sec:methods}
We use 3-D global climate model (GCM) results of a specific H\textsubscript{2}O-rich atmosphere case from \citet{2017APJRKK} as the input for our 1-D models in this study (see Table \ref{table:physicalinput}, more details in Section \ref{sec:3D}). Specifically, we use terminator mean vertical profiles of P, T, N\textsubscript{2} and H\textsubscript{2}O from the GCM as inputs for a 1-D photochemistry model. We augment our 1-D atmospheres with other species (see Section \ref{sec:1D}), including a "modern Earth" constant CO\textsubscript{2} mixing ratio of 360 ppm. We let the atmosphere evolve for five different UV cases (see Section \ref{sec:UV}), including the original UV-quiet model star used in the \citet{2017APJRKK} study. We determine steady state abundances for all modeled species for each simulated atmosphere. We then generate transmission spectra to determine the spectral observables of these habitable moist greenhouse atmospheres with self-consistent photochemistry.
%\vspace{-0.2cm}
\begin{table}
\centering
\caption{\label{table:physicalinput} 1-D Model Parameters: Planet and Star Properties}
\renewcommand{\arraystretch}{1.0}
\begin{tabular}{lc}
\hline
\hline
Parameter                  & Value                                      \\
\hline
Planet Mass                & 1\mearth                                   \\
Planet Radius              & 1\rearth                                   \\
Planet Surface Gravity\tablenotemark{1}     & 1\emph{g}                                  \\
Planet Surface Pressure\tablenotemark{1}    & 1.008 bar                                \\
Planet Surface Temperature & 266 K                                   \\
Star Radius                & 0.137\rsun                                 \\
Star Temperature \emph{T}\textsubscript{eff} & 3300 K (Spectral Type M3)  \\
Star Insolation Flux       & 1650 W/m$^2$ (1.21\emph{S\textsubscript{o}})        \\
%Solar Zenith Angle         & 50 degrees (substellar value)                \\
\hline
\end{tabular}
\tablenotetext{1}{Surface values are terminator mean values, not global mean.}
\vspace{-0.2cm}
\end{table}

The 3-D and 1-D models are not coupled--the GCM is run first and the output fed to the photochemical model, but the output from the latter does not provide feedback into the GCM. Currently, there is no 3-D model available with full chemical mapping capabilities for planets around M dwarfs.  GCMs also do not extend to very low pressures, and thus cannot fully capture dynamics affecting upper atmosphere composition. The GCM is the Community Atmosphere Model (CAM), developed by the National Center for Atmospheric Research (NCAR) to simulate the climate of Earth \citep{Neale2010}. For the \citet{2017APJRKK} work, Version 4 (CAM4) was adapted to simulate Earth-like aquaplanets (i.e. an ocean planet with no land) around M dwarf stars. For our 1-D chemical modeling, we use the \texttt{Atmos} 1-D photochemical modeling tool, most recently developed by the Virtual Planetary Laboratory (VPL). \texttt{Atmos} comes with stellar flux data from the VPL spectral database. We use this data and stellar spectra from the MUSCLES Treasury Survey \citep{2016ApJYoungblood} to provide a range of UV fluxes for the UV-active simulations. This way we are able to realistically portray time-averaged low, medium, and high UV activity. We run a single 1-D model per UV case, for a total of five UV cases (Inactive/Low UV, Medium UV 1, Medium UV 2, High UV, Very High UV). We thus run five photochemical models for the same planet (see Table \ref{table:physicalinput}), each initiated with same physical properties and mixing ratio profiles. We use the \texttt{Exo-Transmit} spectral tool to compute the transmission spectrum for each UV model \citep{2017PASPEK}.

\subsection{Planet Parameters: Moist Atmosphere Simulations}
\label{sec:3D}
CAM has been widely used for studies of habitable exoplanets
\citep{2013APJYang,2014ApJYang,2016ApJYang, 2013AsBioShields, 2014APJHuYang, 2014GeoRLWolfToon, 2015JGRDWolfToon, 2016APJRKK, 2017APJRKK, 2017ApJWolf,2018APJHMET,2014ApJWang,2016ApJWang,2018EPSLBin}. CAM4 has been updated to include a flexible correlated-k radiative transfer module \citep{2013AsBioWolfToon}, and updated water-vapor absorption coefficients from HITRAN2012. The moist habitable atmospheres computed by CAM4 \citet{2017APJRKK} assumed an Earth mass aquaplanet, with a 50m thick "slab" ocean. The slab ocean acts as a thermodynamic layer, where energy fluxes (radiant, latent, and sensible) are calculated between atmosphere and ocean. There is no ocean heat transport, thus the temperature of the ocean is set by surface energy exchange process only. All simulations assumed an atmosphere composed of 1 bar of \molnit{} and variable \water{}. Further details on the numerical scheme can be found in \citet{Neale2010} and Section 2.1 of \citet{2017APJRKK}.\looseness=-1

%The horizontal resolution is 4x5 degrees, and 40 vertical levels are used with an upper boundary at \texttt{\char`\~}1 mbar. 

We use the terminator mean surface values and vertical profiles from a single GCM run from \citet{2017APJRKK} as the inputs for our 1-D models here. We use the planetary terminator results of the 3300 K model star irradiating the synchronously rotating (orbital-rotational period: 19.5724 days) Earth-like planet (see Table \ref{table:physicalinput}) at 1650 W/m\textsuperscript{2} (1.21\emph{S\textsubscript{o}}).

In \citet{2017APJRKK}, GCM simulations were conducted for HZ planets around six different M dwarf model stars (\emph{T}\textsubscript{eff} = 2600 K, 3000 K 3300 K, 3700 K, 4000 K, 4500 K).  The study focused on climate states near the inner edge of the HZ, the runaway greenhouse effect, and stratospheric H\textsubscript{2}O.  Thus the model planets were subject to increasing stellar fluxes until a runaway greenhouse was triggered, marking the terminal inner edge of the HZ. The runaway greenhouse effect is characterized by a collapse of the substellar cloud deck, and a sharp reduction in the planet's albedo.  Temperatures rise rapidly, as a large top of atmosphere (TOA) energy imbalance is maintained, and the increasingly strong water vapor greenhouse prevents the planet from radiatively cooling (e.g. Figures 6 and 7 from \citet{2017APJRKK}). 

%As the atmospheres transitioned to runaway greenhouse climates from stable mild climates for the three stars with \emph{T}\textsubscript{eff} $\leq$ 3000 K, we only consider the coolest \emph{T}\textsubscript{eff} star that generated stable moist greenhouse atmospheres for our study: the 3300 K star. While the 3700 K, 4000 K, and 4500 K star GCM results also feature stable moist greenhouse regimes, detections around a 3300 K star should be the easiest as a smaller star means 1) the planet transit signals are larger, and 2) shorter orbital periods, and thus greater chance of detection with {\TESS} for instance. The 3300 K star was thereby the sole case for which spectra was shown and implications for MIRI observations discussed in \citet{2017APJRKK}. So we are able to compare to those results directly in Section \ref{sec:discussion}. 

For planets orbiting M dwarf stars with \emph{T}\textsubscript{eff} $\leq$ 3000 K, atmospheres transitioned from mild climates with little stratospheric water vapor directly to a runaway greenhouse, with no stable moist greenhouse state existing between (but see \citet{2018EPSLBin}).  For stars with \emph{T}\textsubscript{eff} $\geq$ 3300 K, \citet{2017APJRKK} found that the planets can maintain a stable moist greenhouse regime with their climate remaining stable against a runaway greenhouse.  For our study here, we choose a stable moist greenhouse state around a 3300 K star.  Detections around a 3300 K star will be easier than those around larger hosts because a smaller star means larger transit signals, and a shorter orbital period means that more transit observations can be stacked together in a shorter period of time, improving signal to noise.  The 3300 K star was thereby the sole case for which spectra were shown and implications for MIRI observations discussed in \citet{2017APJRKK}. So we are able to compare to those results directly in Section \ref{sec:discussion}.

Our input P-T profile (i.e. mean terminator thermal profile from the GCM) shows a temperature minimum at 2 mbar, which translates to a cold trap in the H\textsubscript{2}O profile where cloud condensation would occur. Table 1 of \citet{2017APJRKK} reports a GCM model top H\textsubscript{2}O mixing ratio of 5.55x10\textsuperscript{-4}. While this is the global mean value across stratosphere, the mean terminator value is similar due to the strong mixing. Since photolysis of water vapor increases with altitude due to the strengthening incoming stellar UV, we also want to explore the region above 1 mbar in this study. Thus we extend the 1 mbar H\textsubscript{2}O and $T(P)$ values to a TOA pressure of 8.1x10\textsuperscript{-7} bar in our 1-D simulations by simply holding them constant at those values for the atmosphere above 1 mbar in the input profile; photochemical kinetics drives the ultimate steady state abundances.

\subsection{Stellar Parameters: Variable UV Activity}
\label{sec:UV}
\begin{figure}
%\plotone{Binned3300K-1E7.eps}
 %\includegraphics[width=87mm,scale=1.0]{Binned3300K.eps}
 \subfloat[]{%
  \includegraphics[clip,width=\columnwidth,trim=0cm 1.2cm 0cm 0cm, clip=true]{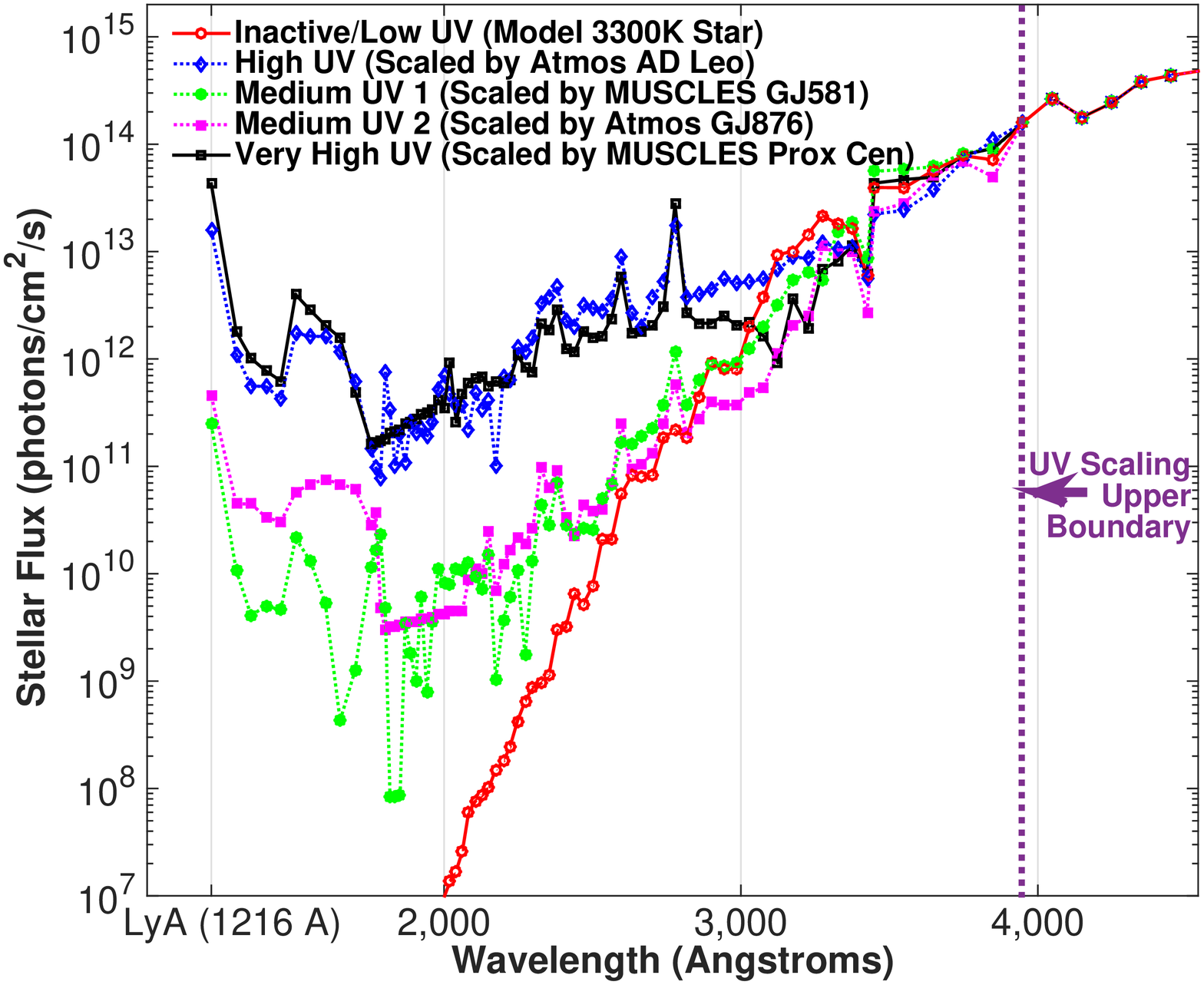}%
} \\[-5ex]

\subfloat[]{%
  \includegraphics[clip,width=\columnwidth,trim=2.5mm 0cm 0cm 0mm, clip=true]{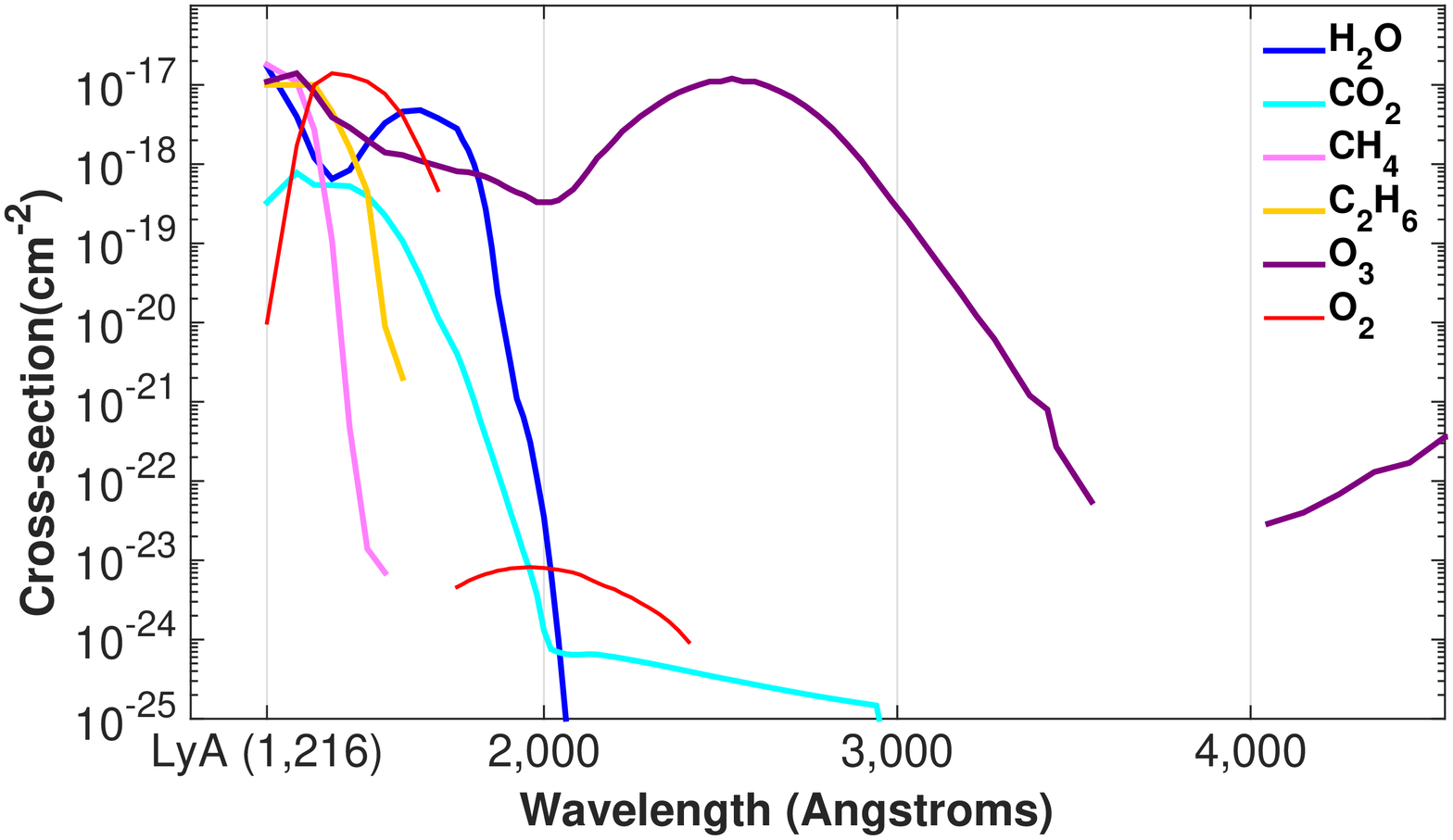}
} \\[-3ex]
\caption{\emph{Top Panel:} Earth-equivalent spectral energy distribution is shown for the inactive model star, with the four synthetically constructed < 3950 {\AA} profiles overlaid. The spectra result from binning the original high resolution data over the \texttt{Atmos} wavelength grid for the UV cross-section data, after converting the flux density units from ergs/cm\textsuperscript{2}/s/A\textsuperscript{-1} to photons/cm\textsuperscript{2}/s/A\textsuperscript{-1} as expected by \texttt{Atmos}. The UV scaling boundary is marked at 3950 {\AA} by the purple vertical line. Since we use real stellar UV data for the scaling, the difference amongst the profiles is not particularly dramatic after the binning. Earth-equivalent fluxes are obtained by taking the original stellar data and dividing by the solar constant. \emph{Bottom Panel:} Available UV cross-section data used by \texttt{Atmos} shown for the species relevant to our 1-D model results. \label{fig:fig1}}
\vspace{-0.2cm}
\end{figure}

We extract high-resolution stellar data of a 3300K UV-quiet star from 1000 to 8500 {\AA}, from the BT-Settl grid of models \citep{2003IAUAllard,2007A&AAllard}. This model spectrum was used in the \citet{2017APJRKK} work; it is our starting scenario---the lowest activity boundary exemplifying a no stellar activity end-member case, with UV from blackbody emission only (red spectrum in Figure \ref{fig:fig1}). However, real stars are not perfect blackbodies at UV wavelengths like the BT-Settl model stars. All stars, even the oldest ones, show some level of chromospheric emission that adds to the UV spectra. 

We incorporate the full spectra (1216-8450 {\AA}) into \texttt{Atmos}, but note that the UV region primarily drives the photochemistry. In addition to the BT-Settl model, we use high-resolution spectra of the M dwarf stars from the MUSCLES database. We choose two stars with the most divergent wavelength-dependent UV (GJ581 and Proxima Centauri)--from the seven M dwarfs in this database--to generate two of the four other UV scenarios. For the remaining two cases, we use AD Leo and GJ876 data from the existing \texttt{Atmos} spectral database. Both spectra feature prominent Ly-$\alpha$.

\captionsetup[subfigure]{labelformat=empty}

We create each synthetic UV spectrum by stitching the extracted data to our UV-quiet model flux data, after binning our stellar data to a pre-defined coarser wavelength grid used by default in \texttt{Atmos} (see Figure \ref{fig:fig1} grid). The "stitching together" is done by scaling the stellar data (up to 3950 {\AA}), such that the value at 3950 {\AA} matches the model star value. In sum: shortward of 3950 {\AA}, we use the scaled extracted data; longward of 3950 {\AA}, we use the model star flux.  \texttt{Atmos}' spectral database stores unscaled stellar flux data as Earth-equivalent incident flux values (\emph{S\textsubscript{o}}) (red spectrum in Figure \ref{fig:fig1}). This means that the overall incident flux at the top of the atmosphere is 1.21 times the fluxes shown in Figure \ref{fig:fig1}. 

The AD Leo (spectral type: M3.5V) and GJ876 (M4V) spectra have been used in several recent 1-D simulation studies of planetary atmospheres using our photochemical model \citep{2005AsBioSegura, 2010AsBioSegura, 2014ApJSDDG, 2015ApJHarman,2017ApJArney}. Their scaled versions correspond to "High UV" and "Medium UV 2", respectively, in Figure \ref{fig:fig1}. We assign the highest activity level ("Very High UV") to the scaled MUSCLES Proxima Centauri (M6V) data. The remaining case ("Medium UV1") corresponds to scaled GJ581 (M3V) MUSCLES UV spectra.

\subsection{Photochemistry in the Atmosphere via {\textit{Atmos}}}
\label{sec:1D}
VPL's \texttt{Atmos} has been used in 1-D photochemical and climate modeling of early Mars, and the Archean and modern Earth atmospheres. \texttt{Atmos} simulations of rocky exoplanets have helped define the HZ \citep{2013aAPJRKK,2013bAPJRKK,2014APJRKK}.

The photochemical model in \texttt{Atmos} \citep{2016AsBioArney,2017ApJArney} solves a set of nonlinear, coupled ordinary differential equations for the mixing ratios of all species at all heights using the reverse Euler method. The method is first order in time and uses second-order centered finite differences in space. The system of equations is explicitly formulated as time-dependent equations that are solved implicitly by a time-marching algorithm. The model is run to steady state to obtain the final mixing ratio profiles. In each step, the model measures the relative change of the concentration of each species in each layer of the atmosphere. When all species in all layers change less their concentrations by than 15\% in the time step, the size of the time step grows. When this size is greater than 10\textsuperscript{17} seconds (\texttt{\char`\~}3 billion years) the model is considered to have reached the steady state (i.e. convergence). This means that the steady state solutions have stable chemical profiles on timescales of billions of years, assuming constant boundary conditions. These boundary conditions can include biological gas fluxes, volcanic outgassing, atmospheric escape and parametrization for ocean chemistry.

We use the species and reaction list of the existing "Modern Earth" template\footnote{Public \texttt{Atmos}: \href{https://github.com/VirtualPlanetaryLaboratory/atmos/}{https://github.com/VirtualPlanetaryLaboratory/atmos/}} in \texttt{Atmos}. Note that our use of this template does not imply that we are attempting to reproduce modern Earth's atmosphere; this template is used for its representative mixture of gases for a \molnit{}-\water{} dominated planet. The base model has 193 forward chemical reactions and 40 photolysis reactions for 40 chemical species made from H, C, O, N, and S, 23 of which participate in photolysis. While CO\textsubscript{2} is kept constant at 360 ppm in all five models, other species are allowed to vary. While a thick CO\textsubscript{2} atmosphere could radiatively cool the middle atmosphere, acting as a bottleneck for H\textsubscript{2}O loss to space, the Earth-like CO\textsubscript{2} amount we assume here should have a relatively minor effect on stratospheric temperatures \citep{2013ApJW&P}.%\looseness=-1

We modify the boundary conditions to simulate an abiotic planet after \citet{2015ApJHarman}. In all five runs, we fix the surface flux of CH\textsubscript{4} to a 1-Earth mass planet abiotic production rate of 1x10\textsuperscript{8} molecules/cm\textsuperscript{2}/s after \citet{2013AsBioGuzman}. To determine the lower boundary conditions for the other varying species (Table \ref{table:chemicalinput} \emph{cont.}), we assume the surface environment (atmosphere and ocean) obeys redox balance (i.e. free electrons are conserved), using the methodology by \citet{2015ApJHarman}. While the atmospheric model ensures redox balance in the atmosphere, we must tune the boundary conditions to prevent a redox imbalance in the oceans or on the planet's surface. We do this in our model by changing the H\textsubscript{2} \emph{flux} into the atmosphere so that there is no net deposition of oxidizing or reducing power into the oceans or onto the surface. Thus, we tune the H\textsubscript{2} \emph{flux} across the five UV scenarios until atmosphere and ocean redox balance independently for each case. This ensures that the atmospheric concentrations we report here are sustainable over geological timescales with only geological (and not biological) fluxes.
%The amount of reducing material flowing out of the ocean into the atmosphere can be controlled by specifying a H\textsubscript{2} surface flux value that balances the amount of oxidants. Thus, we vary the H\textsubscript{2} \emph{flux} across each of the five UV scenarios until global redox balance is achieved for each case. Conserving redox balance ensures that the ocean composition, and thus the atmospheric concentrations we report here, are sustainable over geological timescales with only geological (and not biological) fluxes. 

H\textsubscript{2}O is the only non-background species with mixing ratios provided by the GCM. So, instead of defining surface conditions, we fix the H\textsubscript{2}O abundance profile below the tropopause to the mixing ratios of those levels from the GCM. We assume that the atmosphere is hydrostatic, and there is no atmospheric escape occurring. Photochemical reaction timescales are much shorter (seconds to hundreds of years) than the timescales of atmospheric escape (billions of years). Thus, it is reasonable to exclude the effects of H escape on the chemistry; escape should primarily impact the long term evolution of the atmosphere and not its steady state composition.

%\vspace{-0.1cm}
\begin{table}[htp]
\begin{center}
\caption{\label{table:chemicalinput}Some key photochemical model species with boundary conditions}
\renewcommand{\arraystretch}{1.0}
\begin{tabular}{lll}
\hline
\hline
Species  & Lower bound type  & Lower bound value         \\
\hline
O   & $\nu$\textsubscript{dep} &  1.\\
O\textsubscript{2} & $\nu$\textsubscript{dep} & 1.5$\times$10\textsuperscript{-4} \\
H\textsubscript{2}O  & see table comment & from GCM result \\
H  & $\nu$\textsubscript{dep} &  1.\\
OH & $\nu$\textsubscript{dep}  &  1.  \\
HO\textsubscript{2}  & $\nu$\textsubscript{dep} &  1.  \\
H\textsubscript{2}O\textsubscript{2}  & $\nu$\textsubscript{dep} & 2.$\times$ 10\textsuperscript{-1}  \\
H\textsubscript{2}  & $\nu$\textsubscript{dep}, \emph{flux}, \emph{disth}  &  2.4$\times$10\textsuperscript{-4}, 4.$\times$10\textsuperscript{5}- 2.$\times$10\textsuperscript{10}, 1.\looseness=-1 \\
CO    &  $\nu$\textsubscript{dep} & 1.$\times$10\textsuperscript{-8}\\
CO\textsubscript{2}   &  \emph{fCO\textsubscript{2}} & 3.6$\times$10\textsuperscript{-4}\\
HCO   &  $\nu$\textsubscript{dep} &  1.\\
H\textsubscript{2}CO  &  $\nu$\textsubscript{dep} & 2.$\times$10\textsuperscript{-1}  \\
CH\textsubscript{4}   & \emph{flux} & 1.$\times$10\textsuperscript{8}\\
CH\textsubscript{3}   &  $\nu$\textsubscript{dep} & 1.  \\
C\textsubscript{2}H\textsubscript{6}  &  $\nu$\textsubscript{dep} &  0. \\
NO    &  $\nu$\textsubscript{dep} & 3.$\times$10\textsuperscript{-4} \\
NO\textsubscript{2}   &  $\nu$\textsubscript{dep} & 3.$\times$10\textsuperscript{-3}  \\
HNO   &  $\nu$\textsubscript{dep} & 1. \\
O\textsubscript{3}    &  $\nu$\textsubscript{dep} & 7.$\times$10\textsuperscript{-2}  \\
HNO\textsubscript{3}  &  $\nu$\textsubscript{dep} & 2.$\times$10\textsuperscript{-1}  \\
N  &  $\nu$\textsubscript{dep} & 0.  \\
NO\textsubscript{3}   &  $\nu$\textsubscript{dep} & 0.  \\
N\textsubscript{2}O   &  $\nu$\textsubscript{dep} & 0.  \\
HO\textsubscript{2}NO\textsubscript{2} &  $\nu$\textsubscript{dep} & 2.$\times$10\textsuperscript{-1}  \\
N\textsubscript{2}O\textsubscript{5}  &  $\nu$\textsubscript{dep} & 0.  \\
N\textsubscript{2}  &  \emph{fixedN\textsubscript{2}} & 0.9956 \\
\hline
\end{tabular}
\end{center}
\vspace{-0.4cm}
\tablenotetext{}{\hspace{-0.2cm}\textbf{Note:}  Starting boundary conditions are fixed surface deposition efficiency ("$\nu$\textsubscript{dep}"), constant mixing ratio ("\emph{fCO\textsubscript{2}}"), fixed mixing ratio at the surface ("\emph{fixedN\textsubscript{2}}"), or constant upward flux ("\emph{flux}"); first three quantities are dimensionless, fluxes are in molecules/cm\textsuperscript{2}/s. H\textsubscript{2}O concentration below the tropopause is held at the input H\textsubscript{2}O values from the GCM. H\textsubscript{2} is defined by both $\nu$\textsubscript{dep} and a vertically distributed upward \emph{flux} over a height of \emph{disth} (in km). A range is given for H\textsubscript{2} \emph{flux} as it is the only condition allowed to vary across the five cases to ensure redox balance in the oceans \citep{2015ApJHarman}. S-based species are not shown in this list. While these species (H\textsubscript{2}S, HS, S, SO, SO\textsubscript{2}, H\textsubscript{2}SO\textsubscript{4}, HSO, S\textsubscript{2}, S\textsubscript{4}, S\textsubscript{8}, SO\textsubscript{3}, OCS, S\textsubscript{3}, SO\textsubscript{4} and S\textsubscript{8} aerosols) are retained from the validated Modern Earth template's list to assure convergence, extremely low arbitrary boundary values are supplied to keep their presence negligible (mixing ratio < 10\textsuperscript{-30}).}
\end{table}
%\vspace{-0.2cm}

The vertical grid is distributed evenly over 200 levels and extends to a TOA altitude of 91 km. The lower boundary pressure is set at 1 bar and the upper at 8.1x10\textsuperscript{-7} bar. Both chemistry and vertical transport are considered for the 40 long-lived species. Transport is neglected for the 9 additional short-lived species such as O(\textsuperscript{1}D). For the long-lived species, vertical transport is approximated with a profile of eddy diffusion coefficients. Typically, these coefficients are patterned after eddy diffusivities that best reproduce modern-day Earth \citep{1979Kasting,1990Kasting}.  However, the vertical mixing has been shown to be stronger for Earth-like planets with slow rotation rates, owing to the aforementioned strong substellar convection. Thus for our runs, we adopt a constant eddy profile by iteratively determining a single eddy coefficient value (\emph{K\textsubscript{zz}} = 8.95 x 10\textsuperscript{6} cm\textsuperscript{2}s\textsuperscript{-1}) that allows us to maintain the original GCM H\textsubscript{2}O mixing ratio at 1 mbar, while letting the atmosphere column above vary. Above 1 mbar, we expect photochemistry to dominate over the vertical mixing.

\begin{figure*}
\begin{center}
 \includegraphics[width=160mm,scale=1.0,trim=1cm 0cm 0cm 0mm]{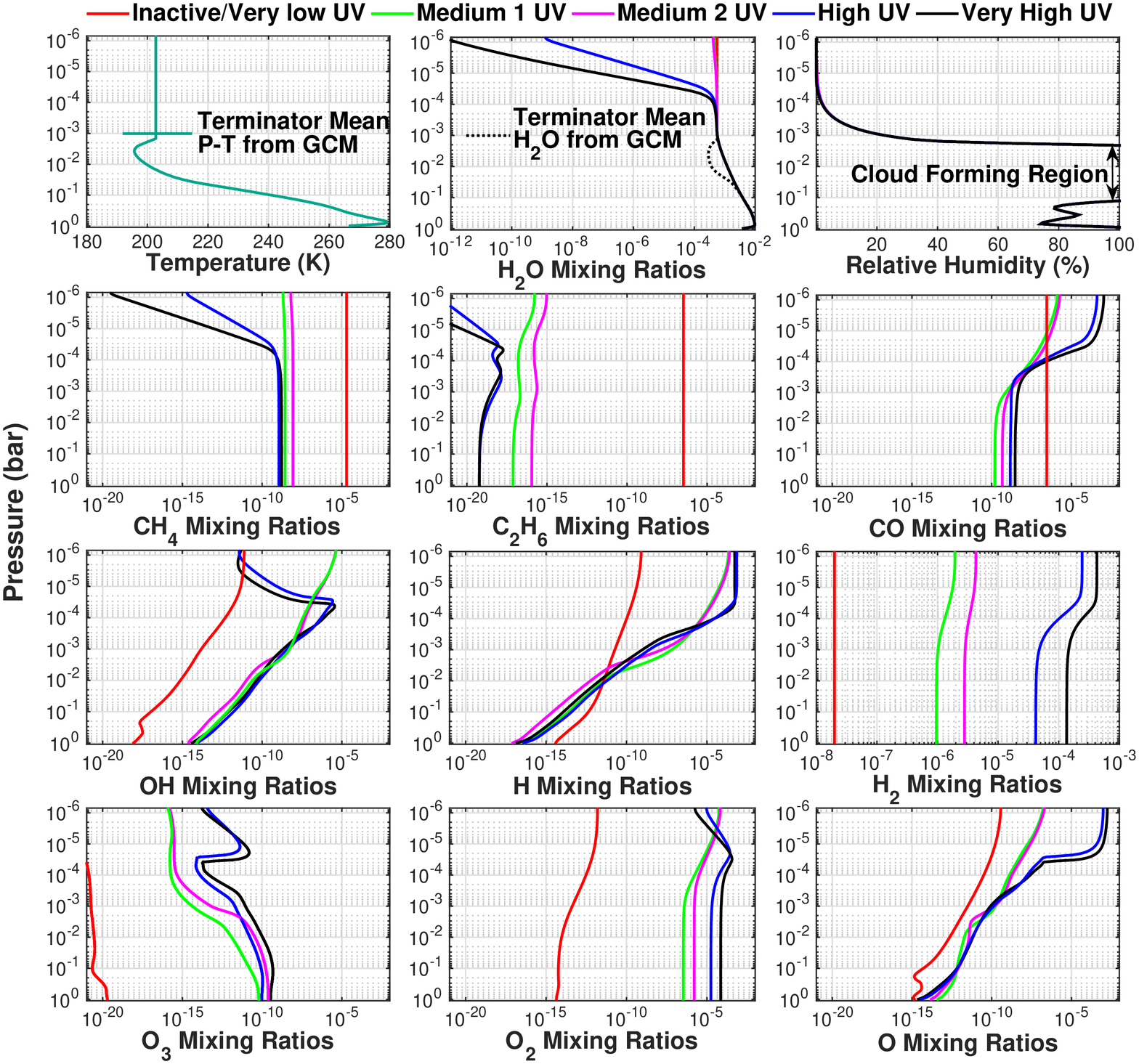}
%\plotone{MixingRatios12Panel_RElH.eps}
\caption{Steady state mixing ratio profiles computed by the photochemical model are shown for all varying non-trace species of interest. 1-D model input for the terminator mean \water{} (dotted-black in the \water{} panel) and $P$-$T$ (dark cyan) profiles from the GCM simulation are shown next to each other for interpretation. With the exception of the H\textsubscript{2} and \water{} panels, all other species are shown on the same wide mixing ratio scale for easy visual comparison. Relativity humidity profiles from the 1-D model for the resulting atmospheres are also shown, with the cloud condensation pressure range indicated within the panel.}
\end{center}
\label{fig:fig2}
\end{figure*}

\subsection{Radiative Transfer Spectra via Exo-Transmit}
\label{sec:exotransmit}
\texttt{Exo-Transmit} is an open source software package to calculate exoplanet transmission spectra. Here we use \texttt{Exo-Transmit} to generate spectra from the computed steady state mixing ratios of all species for which spectral contribution has been established in IR, including trace species for which opacity data is available in the package.

\texttt{Exo-Transmit} is designed to generate spectra of planets with a wide range of atmospheric composition, temperature, surface gravity, size and host star. There is also an option to include an optically thick gray "cloud" deck at a user-specified pressure above the surface. As this cloud deck is not modeled from actual particles, wavelength-dependent cloud properties are not involved. This simply serves as a reasonable method for incorporating the effects of optical thick cloud layers in simulated transmission calculations. When this feature is employed, the transit base is raised to the user-specified cloud-top pressure, meaning data pertaining to the atmosphere column below this pressure level is not read. Our purpose is to quantify spectral differences stemming from varying the UV alone; we keep the transit radius at the model base of 1 bar in our simulations, thus calculating the maximum signal that we would obtain for these atmospheres. We do, however, utilize the cloud truncation feature to compare our spectra to the spectrum shown in \citet{2017APJRKK}, which did include clouds (see Section \ref{sec:discussion}). \texttt{Exo-Transmit} is available publicly on Github with open-source licensing at \href{https://github.com/elizakempton/Exo{\textunderscore}Transmit}{https://github.com/elizakempton/Exo{\textunderscore}Transmit}. 

\texttt{Exo-Transmit} comes with pre-defined P-T and mixing ratio profiles binned to the resolution of its opacity data, where the pressure grid spans 10\textsuperscript{-9} - 10\textsuperscript{3} bar in logarithmic steps of one dex (i.e. $P$ = 10\textsuperscript{$n$}, where $n$ is an integer). For our study, we replace these input files with our own ones containing the newly computed mixing ratios and the mean terminator P-T profile. Our P-T profile is much more finely sampled than \texttt{Exo-Transmit}'s opacity/chemistry grid. During each run of \texttt{Exo-Transmit}, the opacity is first interpolated onto each point of our P-T grid, then the radiative transfer calculation is run to compute the net spectrum.

\section{Results}
\label{sec:results}
\subsection{Atmospheric Constituent Mixing Ratio Profiles}
\begin{figure*}
\begin{center}
\includegraphics[width=154mm,scale=1.0,trim=1cm 0cm 0cm 0mm]{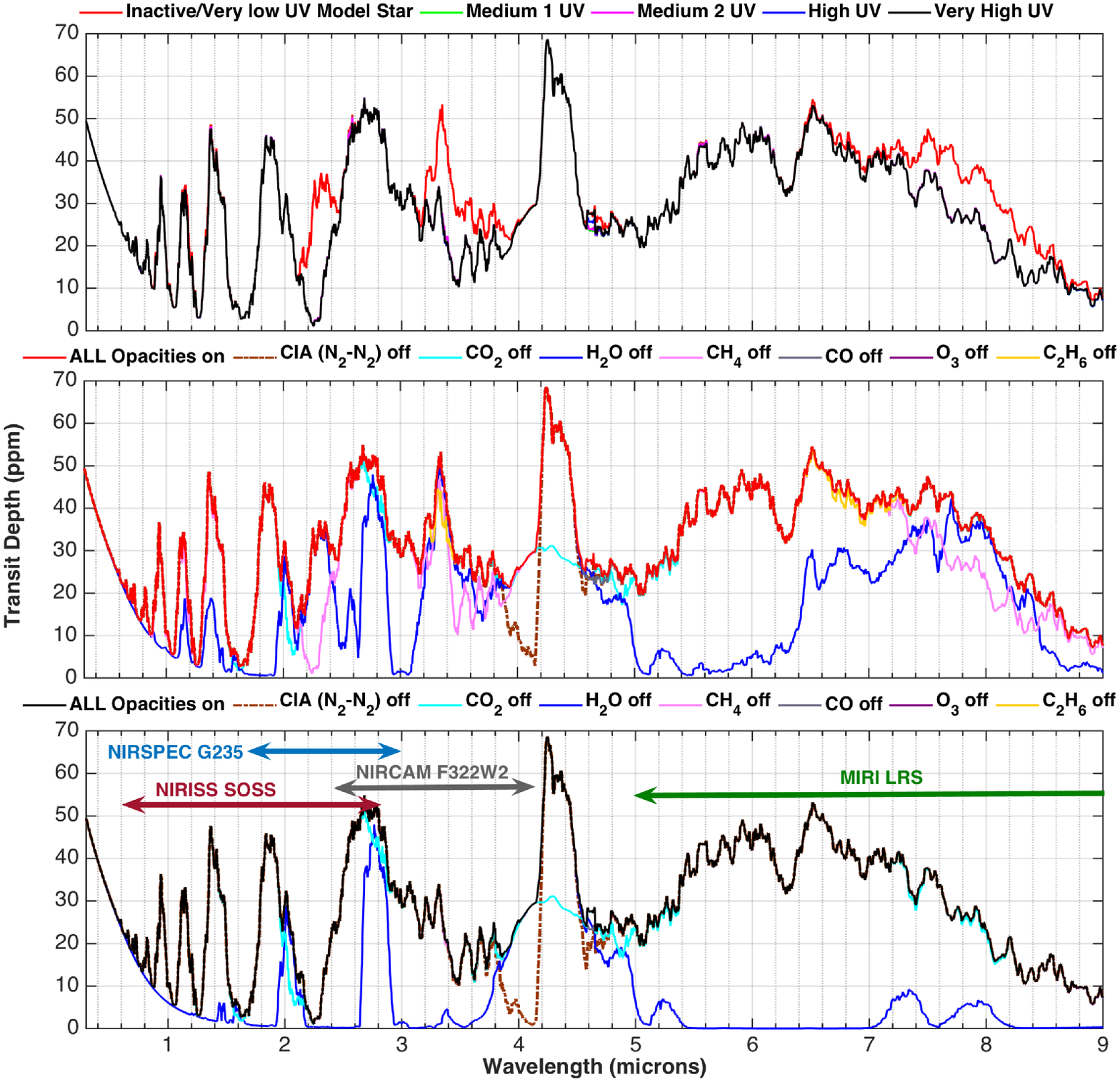}
%\plotone{Fig3rem9umhorlegJWSTFilters.eps}
\caption{\emph{Top panel:} Full spectra from \texttt{Exo-Transmit} shown for all five UV scenarios. Spectra assume there are no clouds. The four synthetic UV active cases completely overlap, with a slight mismatch at 4.3 {\textmu}m due to CO differences. \emph{Middle panel:} Full spectra from the inactive model star case shown (red, per Figure \ref{fig:fig2}'s convention), along with individual contributions removed. \emph{Bottom panel:} Full spectra for the Very High UV case (black, also same convention) shown this time, along with the same individual species removed. The \JWST{} instrument bandpasses where prominent H\textsubscript{2}O features occur have also been indicated here.} 
\end{center}
\label{fig:fig3}
\end{figure*}

Although the two Medium UV spectra have much higher UV levels than the inactive model spectra at wavelengths < 3950{\AA}, with \texttt{\char`\~}10 orders of magnitude difference at 1216 {\AA} (Figure \ref{fig:fig1}), we find neither of them to be high enough to cause appreciable \water{} loss (see top of \water{} panel in Figure \ref{fig:fig2}).

In the inactive model star case--where no photolytic loss is noted for any of the species shown--we see high constant values for the \methane{} and \ethane{} mixing ratios being maintained over the entire atmosphere. The \methane{} source in this abiotic atmosphere is dominated by the lower boundary constant upward flux (\emph{flux} = 1$\times$10\textsuperscript{8} molecules/cm\textsuperscript{2}/s in Table \ref{table:chemicalinput}). So we can expect total \methane{} production to be similar across all five cases. For the four UV-active cases, higher UV would translate to higher total loss of \methane{} due to photolysis. \methyl{} is a highly reactive product of CH\textsubscript{4} photolysis and combines with itself to produce \ethane{}, which also photodisassociates and cycles back into \methane{}; the three hydrocarbons cycle amongst themselves. The dominant exit pathway from this cycle - reaction of \methyl{} with oxidants - is limited by the rate of \methane{} + OH ---> \methyl{} + \water{}. Thus, \methane{} chemistry is dominated by the abundance of OH in the atmosphere, as evidenced by a negative correlation between the OH and \methane{} profiles. The higher the OH abundance, the higher the \methane{} loss via combination with OH. This is also demonstrated by our examination of the sinks for \methane{}; for UV-active stars, destruction of \methane{} is dominated by reaction with OH, and this rate even outpaces our abiotic surface \methane{} flux. The net impact of this chemistry is much lower abundances for \methane{} and \ethane{} for the UV-active stars compared to the inactive model star. 

%\ethane{} mixing ratio profiles follow a more linear trend for the UV-active cases (i.e. higher UV flux leads to less \ethane{}) compared to \methane, as the \ethane{} cross-section profile peaks at the Ly-$\alpha$ wavelength. The similarities between \methane{} and \ethane{} profile trends can be explained by the fact that the three most dominant \methane{} sink reactions all produce \methyl{}, some of which becomes C\textsubscript{2}H\textsubscript{6}.

OH is primarily produced in the atmosphere via \water{} photolysis. Our Medium UV stars show similar Ly-$\alpha$ strengths, so the resultant \water{} photolysis rates are also similar. For the High UV and Very High UV stars--both with synthetic UV \texttt{\char`\~}2 orders of magnitude higher than the Medium UV stars--we see significant H\textsubscript{2}O loss: the High UV star depletes TOA \water{} by \texttt{\char`\~}6 orders of magnitude; while the Very High UV star depletes the TOA \water{} by \texttt{\char`\~}3 further orders of magnitude. Ly-$\alpha$ strength is slightly higher in the Very High UV spectra.\looseness=-1

Like most species in our model, atomic O, O\textsubscript{2}, and O\textsubscript{3}, have only been defined by lower boundary deposition efficiency values. Without a surface source of O\textsubscript{2} and \ozone{}, atmospheric chemistry is their only source. Thus, their concentrations are minor for the inactive star (red profiles in Figure \ref{fig:fig2}). As UV increases, we see higher abundances in all three species, especially \moloxy{} and \ozone{}, although their presence remains below detectable levels. At high altitudes, both O and O\textsubscript{2} generally increase in concentration, consistent with having a photolytic source. O\textsubscript{2} maintains high abundance throughout the column for the highest UV cases as the UV flux is quite high in both spectra between 1300-1500 {\AA}, where the O\textsubscript{2} UV cross-section curve peaks. Although far UV activity is important for O\textsubscript{3} production, O\textsubscript{3} remains minor throughout the atmosphere even for the highest UV case. O\textsubscript{3} is produced in the model by the three-body reaction O + O\textsubscript{2} ---> O\textsubscript{3} and this is the only mechanism that produces it in this atmosphere. Otherwise, O\textsubscript{3} is being destroyed via photolysis reactions O\textsubscript{3} + {\it{h}}$\nu$ ---> O\textsubscript{2} + O(\textsuperscript{1}D) and O\textsubscript{3} + {\it{h}}$\nu$ ---> O\textsubscript{2} + O, along with 10 other chemical reactions. When O\textsubscript{3} is destroyed via these pathways, O\textsubscript{2} is always produced. 
%\ozone{} is the only important species we model for which \emph{Atmos} further computes temperature-dependent cross-sections over the full stellar data range ( i.e. 1216-8450 {\AA) within the model. Therefore, the cross-sections for \ozone{} are altitude-dependent and in in Figure \ref{fig:fig1}, we show the values of the level at which all three alltropes show mixing ratio turning points (3$\times$10\textsuperscript{-5} bar, 69 km). \looseness=-1

\subsection{Transmission Spectra}
For all cases considered, with no cloud cover, we obtain large H\textsubscript{2}O features--around 50 ppm in strength--between 2.5-3.8 {\textmu}m and 4.5-9 {\textmu}m (Figure \ref{fig:fig3}). The locations where these H\textsubscript{2}O features peak are consistent with Figure 11 of \citet{2017APJRKK} (Figure \ref{fig:fig4}). The 2.5-3.8 {\textmu}m feature peaks within the \JWST{} NIRCam grism with the F322W2 filter, as well as the NIRSPEC G235M/H and NIRISS SOSS (Single Object Slitless Spectroscopy) bandpasses. The 4.5-9  {\textmu}m feature is well within the \JWST{} MIRI LRS (Low Resolution Spectroscopy) bandpass. For the model star case, we also see additional absorption at 2.2 {\textmu}m, 3.3 {\textmu}m and 7.4-8.4 {\textmu}m (i.e. the red profile minus the black profile in the top panel of Figure \ref{fig:fig3}), caused by the high \methane{} (>10\textsuperscript{-5}) predicted by the model for our choice of \methane{} lower boundary condition.

CO\textsubscript{2} produces the tallest feature in our spectra (Figure \ref{fig:fig3}). This narrow feature at 4.3 {\textmu}m (\texttt{\char`\~}0.4 {\textmu}m wide when measured from the brown dotted spectrum) is \texttt{\char`\~}20 ppm larger than the H\textsubscript{2}O features, and 40 ppm larger than the N\textsubscript{2}-N\textsubscript{2} collision induced absorption (CIA) feature it overlaps. This region falls within the bandpasses of the NIRCAM F444W filter (3.8-4.8 {\textmu}m) as well as the NIRSPEC G333/H grating (2.9-5 {\textmu}m).

The N\textsubscript{2}-N\textsubscript{2} CIA feature spans the 3.8 to 5 {\textmu}m wavelength range (cyan colored bumps in Figure \ref{fig:fig3}).  This opacity source stems from the deep atmosphere (right panel of Figure \ref{fig:fig4}) and is strong due to the high N\textsubscript{2} \citep{2015ApJSchwieterman} content of the atmosphere. This implies we may be able to quantify atmospheric N\textsubscript{2} from this feature's strength in a cloudless atmosphere. As evident from the two spectra shown with CIA removed (brown dotted plots in Figure \ref{fig:fig3}), the base of the 4.3 {\textmu}m CO\textsubscript{2} feature is broadened by the CIA feature but the height is unaffected; this feature is the only other absorption feature occurring between 4 {\textmu}m and the CO\textsubscript{2} feature.

While O\textsubscript{2} is significantly enhanced with increasing UV, it is spectrally insignificant. The few features from O\textsubscript{2} molecular and collision induced opacities occur < 1.0 {\textmu}m and are completely masked by the Rayleigh tail. Atomic O is enhanced at high altitudes, but there is no atomic opacity for O. O\textsubscript{3} can be spectrally important in IR as it produces molecular and collision induced features at wavelengths longer than 1.1 {\textmu}m. O\textsubscript{3} acts as a UV shield for atmospheres where it is present (e.g. Earth) and thus has important consequences for habitability \citep{2015aAPJRugheimer,2015bAPJRugheimer}.  O\textsubscript{3} appears to have no impact on any of the spectra we have presented here.

Most importantly, Figure \ref{fig:fig3} (top panel) shows that spectra from the UV active runs overlap almost completely. However, in the inactive model star's case, additional absorptions longward of the peak H\textsubscript{2}O features stem primarily from the high \methane{} and some from \ethane{}. \ethane{} contribution is most noteworthy at 3.3 {\textmu}m, where it also overlaps with major contribution from CH\textsubscript{4} and H\textsubscript{2}O, and then from 6.5-7.4 {\textmu}m, where most of the contribution is still from H\textsubscript{2}O. These demonstrate the impact of M dwarf UV activity on the planet's transit observations. We find that for any realistic levels of UV radiation from an M dwarf host star, the atmospheric response to the amount of radiation is undetectable.

From our 1-D modeling results, we infer that H\textsubscript{2}O photodissociation from high UV instellation does not impact transmission spectra, and thus should not affect our ability to detect H\textsubscript{2}O absorption features from a habitable synchronously rotating Earth-like planet around a nearby M dwarf. This is expected as transmission features should primarily originate from lower regions of the atmosphere $\geq$ 1 mbar (left panel of Figure \ref{fig:fig4}). In the upper atmosphere above the substellar clouds, there is no reason for the planet-wide vertical mixing driven by strong convection to continue dominating. Thus photochemistry can dominate here, meaning we cannot expect H\textsubscript{2}O to stay enriched. Accordingly, we only note visible impact from photochemistry above the stratosphere (and cloud decks) for all species involved in photolysis in the modeled atmospheres (Figure \ref{fig:fig2}).\looseness=-1

\section{Discussion}
\label{sec:discussion}
\begin{figure*}
 \includegraphics[width=180mm,scale=1.0]{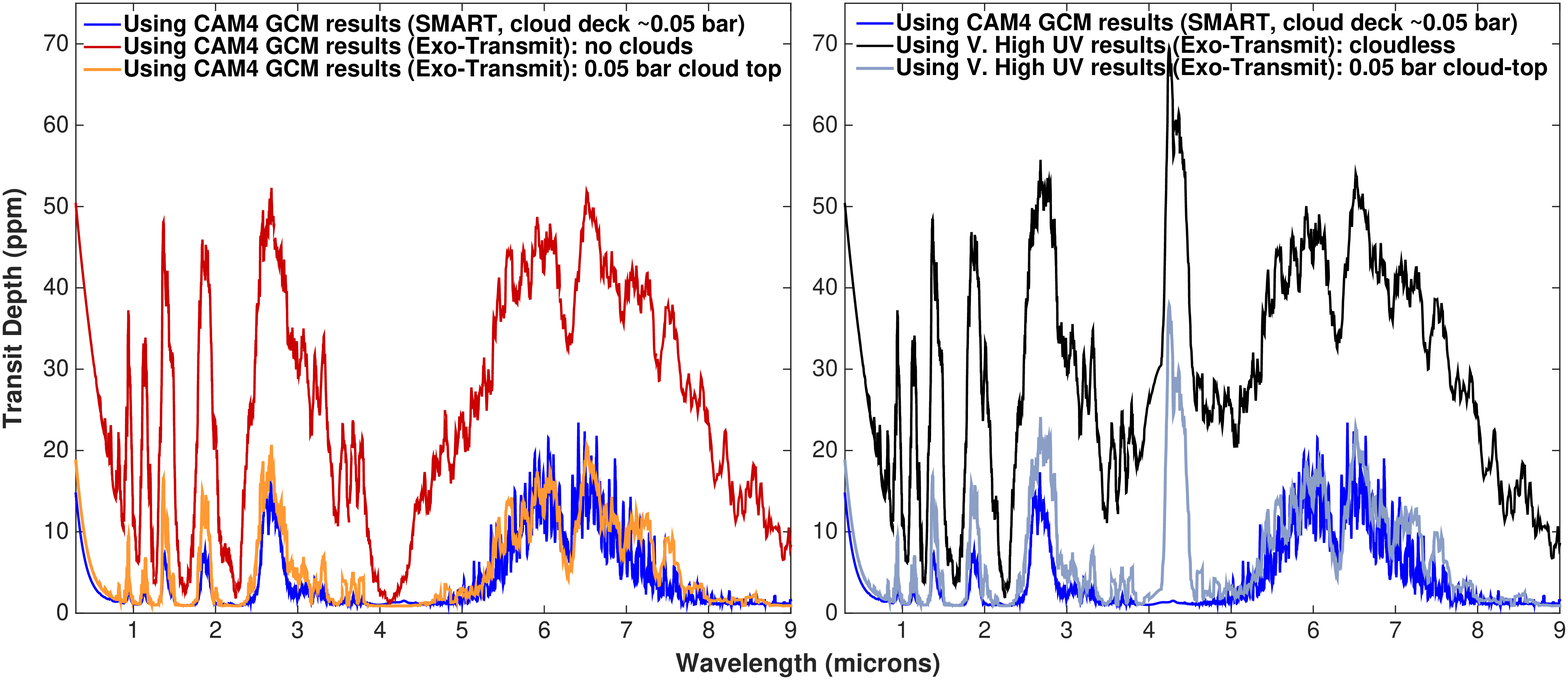}
%\plotone{Fig4ET-updated_0.05barclouds.eps}
\caption{\emph{Left panel:} \texttt{Exo-Transmit} spectra computed using terminator mean mixing ratios from the pure N\textsubscript{2}-H\textsubscript{2}O atmosphere (GCM results) shown alongside the \citet{2017APJRKK} \texttt{SMART} version of it (blue). As the water clouds occur \texttt{\char`\~}0.05 bar in the \texttt{SMART} computations, we also include the spectrum that results from placing an opaque cloud at 0.05 bar within \texttt{Exo-Transmit}. \texttt{Exo-Transmit} spectra shown here have CIA turned off for direct comparison; the \texttt{SMART} spectrum does not include CIA. The CIA opacity has a minimal effect on these spectra so their inclusion or non-inclusion is not a dominant factor in the modeling outcomes. \emph{Right panel:} Very High UV \texttt{Exo-Transmit} spectra shown for cloudless (same black spectrum from Figure \ref{fig:fig3}) and 0.05 bar cloud top (gray) cases, in comparison to the \texttt{SMART} spectrum of the GCM results. Note that the difference in the 4-5 um range between the GCM + \texttt{SMART} simulation (in blue) and the GCM + \texttt{PHOTOCHEM} + \texttt{Exo-transmit} simulations (gray/black spectrum) is due to the inclusion of \carbdiox{} in the photochemical models, but not in the GCM, which assumed a pure \molnit{}-\water{} atmosphere.
\label{fig:fig4}}
\end{figure*}
Figure \ref{fig:fig2} shows that water vapor concentration is highest between 0.1 and 1 bar, close to the surface as expected. The cold trap minimum is also found between two other \texttt{Exo-Transmit} pressure grid points. To ensure accuracy of our interpolations onto this coarse pressure grid, we have experimented with shifting the cold trap altitude. Of-course, the transit depths shown in Figure \ref{fig:fig3} should still not be taken at face value; moist atmospheres have water clouds that are not accounted for here. Since \texttt{Exo-Transmit} does not presently have the capability to include specific cloud properties, we do not include the liquid and ice data from the GCM run outputs as part of our spectral analysis. We see the impact of such water clouds in the transmission spectra provided in \protect{\citet{2017APJRKK}}, computed by VPL's Spectral Mapping Atmospheric Radiative Transfer (\texttt{SMART}) tool for the planet-star configuration we have studied here. Data from the GCM simulation results were used directly in \texttt{SMART} to calculate the spectrum for the synchronously rotating planet.

We are able to compare our spectra with the \texttt{SMART} spectrum by raising the base altitude and thus sampling a lower pressure range (i.e. a shallower column) of the atmosphere in \texttt{Exo-Transmit}. For comparison, we compute spectrum for the original pure \molnit{}-\water{} GCM atmosphere, with no photochemical alteration. We do this for a few cases with \texttt{Exo-Transmit}: a default case with the original 1 bar base, and some additional computations to simulate the effects of clouds by raising the spectral model base to successively lower pressures. In the left panel of Figure \ref{fig:fig4}, we show the spectrum of the default case with no clouds, and the spectrum for a cloud-top pressure of 0.05 bar--the location of the water clouds in the \texttt{SMART}-computed version of the GCM results. For comparison, we overplot the spectrum from \texttt{SMART}. We see that by artificially truncating our spectra at this cloud-top pressure, we are able to capture most of the impacts of the clouds on the spectrum. 

We validate the consistency of our \texttt{Exo-Transmit} calculations with the \texttt{SMART} spectrum by noting a reasonable overlap between the two cloudy spectra (in both panels). We see that the spectra diverge in the near-IR, for wavelengths shorward of 4 {\textmu}m. The transit depths are consistently larger in the cloudy spectrum computed by \texttt{Exo-Transmit} for wavelengths shorter than 4 {\textmu}m; \texttt{Exo-Transmit} computes a larger apparent size for these planets than \texttt{SMART} in the near-IR. Differences in the treatment of clouds (properties such as particle size, optical constants, etc.) between the two spectral models would be responsible for this kind of discrepancy. As we mentioned earlier in Section \ref{sec:exotransmit}, \texttt{Exo-Transmit} treats its clouds as a fully gray optically thick deck, so cloud properties do not vary with wavelength. However, \texttt{SMART} has wavelength-dependent water cloud properties incorporated in the spectral modeling scheme. We believe the trend we are seeing in Figure \ref{fig:fig4} is a result of this.

Figure \ref{fig:fig4} (right panel) also shows that the CO\textsubscript{2} 4.3 {\textmu}m feature remains prominent in the cloudy spectrum of the photochemical model results. The atmospheric column above 0.05 bar is still being transmitted and contributes to the spectrum. At 0.05 bar, the H\textsubscript{2}O mixing ratio is \texttt{\char`\~}2$\times$10\textsuperscript{-3}. Transit spectroscopy typically sample the planet at \texttt{\char`\~}1 mbar atmospheric pressure, which is also the GCM TOA. The GCM TOA H\textsubscript{2}O mixing ratio is \texttt{\char`\~}5.55$\times$10\textsuperscript{-4} (Figure \ref{fig:fig2}, also see Table 1 of \citet{2017APJRKK}). Since our CO\textsubscript{2} abundance is slightly less than that and also less than \texttt{\char`\~}1/5th of the 0.05 bar H\textsubscript{2}O abundance, the much higher CO\textsubscript{2} transit depth signal is noteworthy for future observing efforts.

\citet{2016ApJGreen} has suggested systematic noise floors of  \texttt{\char`\~}20 ppm for NIRISS/NIRSpec and \texttt{\char`\~}30 ppm for the NIRCam grism, which is not only well below the CO\textsubscript{2} 4.3 {\textmu}m signal strength for the cloudless case, but is also comparable to the difference between this feature and the largest H\textsubscript{2}O features. These observational advantages still hold for the 0.05 bar cloud top spectra. So in a similarly cloudy atmosphere but with modern Earth-like amounts of CO\textsubscript{2}, the CO\textsubscript{2} may be more readily detected. Of-course, the true performance of the instruments will only be known once \JWST{} actually flies. 

%We also compute higher tropospheric O\textsubscript{3} for the Medium UV 2 case than the High UV case, and significantly more tropospheric O\textsubscript{3} than the Medium UV 1 case (Figure \ref{fig:fig2}). The latter trend is seen for \moloxy{} as well as \moloxy{} and \ozone{} cycle back and forth. This is expected as the Medium UV 2 (pink) star generally has more UV for wavelengths < 2500 {\AA} than Medium UV 1 (green) star, where the O\textsubscript{3} cross-section profile peaks. We believe the non-linear trend between Medium UV 2 and High UV is the result of the Medium UV 2 case experiencing the least total CH\textsubscript{4} loss of all four UV-active cases, which would explain its lower atomic H and thus higher available O for increased O\textsubscript{3} production. 

One issue that can impact our quantitative predictions of \moloxy{} and \ozone{} is how \texttt{Atmos} treats \carbdiox{}. \texttt{Atmos} presently accepts CO\textsubscript{2} as a user-specified constant mixing ratio for terrestrial planets, though it allows for CO\textsubscript{2} production and destruction reactions, including photolysis (\carbdiox{} + {\it{h}}$\nu$ ---> \carbmono{} + O and \carbdiox{} + {\it{h}}$\nu$ ---> \carbmono{} + O(\textsuperscript{1}D)). So a caveat here is that the model produces excess CO\textsubscript{2} in order to maintain the constant 360 ppm \carbdiox{} abundance. This overestimates both the upper atmosphere \carbdiox{} abundance and the total O budget. This could be important for the computed O\textsubscript{2} and O\textsubscript{3} mixing ratios. However, since we note photochemical effects only above the column of the atmosphere sampled by transits, a major impact on transit observations is unlikely and this impact is mostly just quantitative. For now, we save discussion of \ozone{} absorption features for a subsequent manuscript focused more on O chemistry.

H\textsubscript{2} and CH\textsubscript{4} are the only non-negligible species in our model defined by lower boundary \emph{flux} values (see Table \ref{table:chemicalinput}) that also dominate their source functions. The higher the stellar UV flux, the higher the H\textsubscript{2} \emph{flux} value needs to be to achieve global redox balance for a given CH\textsubscript{4} flux. H and OH trends across the models are thus affected by the varying H\textsubscript{2} source. We keep the CH\textsubscript{4} flux fixed in this study as small changes to this flux do not have an impact on the concentrations of gases for which we note detectable spectral features in Figures \ref{fig:fig3} and \ref{fig:fig4}. Further investigation of the effects of the choice of \methane{} flux on the abundances and detectability of relatively trace species will be explored in a future paper. 

\section{Conclusions}
\label{sec:conclusions}
We have taken a synchronously orbiting aquaplanet-star pair result from \citet{2017APJRKK} within the stable moist habitable regime, and have conducted a case study on its future detectability via \JWST{}. We have investigated the impact of stellar activity on the detection of terrestrial planets with water-rich stratospheres, as photolysis from high UV activity would continuously destroy the water lofted into the high atmosphere. We have run five photochemical models, each with different wavelength-dependent UV activity, ranging from the inactive model (BT-Settl) star data used in \citet{2017APJRKK} to Proxima Centauri-like levels. We have used stellar fluxes from the VPL spectral database in \texttt{Atmos} and the MUSCLES Treasury Survey to vary the UV levels.

We find that as long as the atmosphere is well-mixed up to 1 mbar, \water{} strengths observed in transit spectra should remain unaffected by UV activity. However, for the inactive model star, transit depths are larger due to contributions from the high \methane{} in the atmosphere for our specified \methane{} surface flux, which assumes Earth-like abiotic sources of \methane{}. Detectable \methane{} is absent in the UV active cases.

CO\textsubscript{2} produces a narrow but large detectable feature at 4.3 {\textmu}m. For our assumed atmospheric CO\textsubscript{2} level of 360 ppm, this feature is about 20 ppm larger than the tallest \water{} features. At the wavelengths where CO\textsubscript{2} does overlap \water{} features, the maximum contribution is also 20 ppm over a very narrow (<1 {\textmu}m broad) region at 2 {\textmu}m (i.e. before the prominent H\textsubscript{2}O features appear) for a cloudless case. We also see broadening of the base of the 4.3 {\textmu}m CO\textsubscript{2} feature due to opacity from N\textsubscript{2}-N\textsubscript{2} collisional pairs. However, upon comparing the two "Very High UV" spectra in the right panel of Figure \ref{fig:fig4}, we see that the N\textsubscript{2}-N\textsubscript{2} CIA feature is a high pressure feature originating from well below cloud decks. Thus, the CIA feature should not contribute to the observed signal from CO\textsubscript{2} in practice.

While UV activity may not impact transit depths at the ppm level, water clouds do. Upon comparing the cloud-containing \texttt{SMART} spectrum from \citet{2017APJRKK} for the synchronously rotating planet modeled by the GCM, with a cloud-free version of it from \texttt{Exo-Transmit}, we find that the 0.05 bar cloud weakens the strongest H\textsubscript{2}O feature at 6 {\textmu}m from 50 ppm---comparable to the postulated MIRI systematic noise floor---to only 15 ppm. This means should the terminator cloud deck occur lower in the atmosphere, at pressures > 0.05 bar, we should require less observing time to make a positive detection. Overall, we see few detectable impacts of photochemistry on the spectra of these worlds.\looseness=-1

\acknowledgments
{We thank contributors and fellow developers of the {\it Atmos} software and the open source Exo-Transmit tool used in this study. This work was carried out at the NASA Goddard Space Flight Center (GSFC) under the Center for Research and Exploration in Space Science and Technology (CRESST) II cooperative agreement between the lead author's institution, University of Maryland College Park, and GSFC (Task 699.05). Goddard affiliates are thankful for support from GSFC Sellers Exoplanet Environments Collaboration (SEEC), which is funded by the NASA Planetary Science Division's Internal Scientist Funding Model. This work was performed as part of the NASA Astrobiology Institute's Virtual Planetary Laboratory, supported by NASA under cooperative agreement NNH05ZDA001C. RKK and ETW gratefully acknowledge funding from NASA Habitable Worlds grant NNX16AB61G. The 3D simulation results used in this study were facilitated through the use of advanced computational, storage, and networking infrastructure provided by the Hyak supercomputer system, supported in part by the University of Washington eScience Institute. This work also utilized the Janus supercomputer, which is supported by the National Science Foundation (award number CNS-0821794) and the University of Colorado at Boulder. ETW thanks NASA Astrobiology Institute CAN7 award NNH13DA017C through participation in the Nexus for Exoplanet System Science (NExSS). SDDG and RKK also acknowledge financial support from the NASA Astrobiology Program via NExSS. Any opinions, findings, and conclusions or recommendations expressed in this material are those of the author(s) and do not necessarily reflect the views of NASA or the National Science Foundation.}

\emph{Software}: Atmos, Exo-Transmit \citep{2017PASPEK}, SMART \citep{1996JGRMeadowsCrisp,1997GeoRLCrisp}, MATLAB.

\bibliography{ms}

\end{document}